# Charge state transition of spectrally stabilized tin-vacancy centers in diamond

Keita Ikeda[1], Yiyang Chen[1], Peng Wang[1], Yoshiyuki Miyamoto[2], Takashi Taniguchi[3], Shinobu Onoda[4], Mutsuko Hatano[1], Takayuki Iwasaki[1,*]

[1]Department of Electrical and Electronic Engineering, School of Engineering, Institute of Science Tokyo, Meguro, 152-8552 Tokyo, Japan

[2]Research Center for Computational Design of Advanced Functional Materials, National Institute of Advanced Industrial Science and Technology, Tsukuba, Ibaraki 305-8568, Japan

[3]Research Center for Materials Nanoarchitectonics, National Institute for Materials Science, 305-0044 Tsukuba, Japan

[4]Takasaki Advanced Radiation Research Institute, National Institutes for Quantum Science and Technology, 1233 Watanuki, Takasaki, 370-1292 Gunma, Japan

[*]Email: iwasaki.t.c5b4@m.isct.ac.jp

**Abstract**

Solid-state quantum emitters are an important platform for quantum information processing. The fabrication of the emitters with stable photon frequency and narrow linewidth is a fundamental issue, and it is essential to understand optical conditions under which the emitter keeps a bright charge state or transitions to a dark state. For these purposes, in this study, we investigate the spectral stability and charge state transition of tin-vacancy (SnV) centers in diamond. The photoluminescence excitation spectra of multiple SnV centers are basically stable over time with nearly transform-limited linewidths under resonant excitation, while simultaneous irradiation of resonant and non-resonant lasers makes spectra from the SnV centers unstable. We find that the instability occurs due to the charge state transition to a dark state. The charge state transition rates are quantitatively investigated depending on the laser powers. Lastly, with first-principle calculations, we model the charge state transition of the SnV center under the laser irradiation.

**Introduction**

Color centers in diamond are a promising building block for quantum information processing[1]. An impurity atom and vacancy form a color center and the corresponding unique energy level. The color center shows fluorescence upon laser irradiation and its electron spin state can be also optically readout. This makes it possible to generate the spin-photon interface, which is essential for generation of remote entanglement[2]. The representative color center in diamond is a pair of a nitrogen atom and a vacancy, forming a nitrogen-vacancy (NV) center[3]. The important demonstrations for quantum network include two-photon interference[4,5], a quantum memory over 1 min utilizing nuclear spins[6], photon state transfer into a nuclear spin[7,8], long-distance remote entanglement[9,10], and entanglement among multiple qubits[11]. However, the NV center suffers from a low fraction of the zero-phonon-line (ZPL) and it is subject to spectral diffusion by external noise. Alternative emerging color centers are composed of group-IV elements in the split vacancy configuration[12,13]. Among them, color centers based on the heavy atoms of tin (Sn) and lead (Pb) leads to more robustness to the effect of phonon in the ground state[14–16], expecting a milli-second spin coherence at Kelvin temperatures, compared with the silicon-vacancy (SiV) and germanium-vacancy (GeV) centers[17,18]. Indeed, the tin-vacancy (SnV) center has been shown to have a spin coherent time of 0.3 ms for all-optical control[19] and 0.65 ms[20] and 1.57 ms[21] for microwave control at 1.7 K. A longer spin coherence time of 10 ms has been also demonstrated at 50 mK using a superconducting wire for the microwave irradiation[22]. Nuclear spins close to a SnV center can be also controlled based on dynamical decoupling techniques[23]. The SnV centers shows optical properties such as single photon emission[14,24], transform-limited photon emission[25,26], nearly identical photon frequencies[27], and two-photon interference based on one SnV center[28]. Furthermore, nanophotonic and open cavity structures including SnV centers have been fabricated[28–33].

Understanding the charge state transition of color centers in a solid-state system, including the color center itself and surrounding environment, is crucial to obtain demanding optical and spin characteristics for quantum applications. The negatively-charged state is currently known as a useful charge state of the SnV center, referred to SnV$^-$. A theoretical calculation suggests other charge states such as neutral state (SnV$^0$) and double negatively-charged state (SnV$^{2-}$)[34]. Hereinafter, the simple expression "SnV" refers to the -1 negatively-charged state. Experimentally, resonant excitation to the SnV stochastically leads to the fluorescence termination, indicating the charge state conversion to a dark state[26,35]. Here, the dark state means the state which has different energy levels from the SnV center, and thus, it is not resonant to the original frequency of a tunable laser, i.e. around 619 nm for the SnV center. A non-resonant laser works for recovery to the bright SnV center. Thus, the non-resonant laser pulse is included at the first of the sequence for spin manipulation[19–22]. A green laser is generally used as the charge stabilization laser. However, the non-resonant laser changes the charge state of surrounding defects, resulting in spectral diffusion/jump of the color center. Recently, spectral diffusion of a SnV center was observed upon the irradiation of the non-resonant green (515 nm) laser after vanishing the fluorescence by the resonant excitation[36]. Other reports[26,35] showed that simultaneous irradiation of the resonant and non-resonant green lasers led to unstable photoluminescence excitation (PLE) spectra, probably due to spectral diffusion or charge state transition. Observation of the instability was further indicated for a pulse sequence, mentioning that the initialization of an emitter was not achieved under the simultaneous irradiation[37]. Accordingly, it is challenging to fabricate stable quantum emitters in a solid-state material and important to gain an insight into the charge state dynamics when the emitter shows unstable fluorescence. In this study, we statistically observe the effect of various conditions of the laser irradiation on the stability of the high-quality SnV centers with low strain and low amount defects in the sample such that a narrow inhomogeneous distribution was obtained[27]. We find that spectral diffusion is largely suppressed for multiple SnV centers even after the non-resonant green pulse for the charge stabilization. Furthermore, we also observe unstable fluorescence upon the simultaneous irradiation of the resonant and non-resonant green lasers similar with the previous reports[26,35], and here reveal that the dark state transition occurs using photons from both the lasers. Finally, we construct a model of the charge state transition based on the first-principles calculations.

**Experimental and calculation methods**

The SnV centers are fabricated in IIa-type single-crystal diamonds (Element Six, Electronic grade) by ion implantation followed by HPHT annealing[27]. Ion implantation is conducted at an acceleration energy of 18 MeV with a fluence of $5\times10^8$ cm$^{-2}$, resulting in a projected depth of 3 μm from the surface. Annealing treatment is performed at 2100°C under 7.7 GPa, which has been shown to generate nearly identical photon frequencies from SnV centers in previous study[27]. For all optical experiments, a home-built confocal system equipped with a cryostat is used (Fig. 1a). The measurements are carried out at a low temperature of ~6 K. A green laser (515 nm, Cobolt) was used for non-resonant excitation. Resonant excitation is performed using a dye tunable laser (Sirah Lasertechnik Matisse 2 DS). The wavelength of the dye laser is monitored by a wavemeter (Highfinesse WS8-30). Both the tunable dye laser and the 515 nm laser are intensity modulated by acousto-optic modulators (AOM, Gooch & Housego). Several fibers are used in the laser incidence system including a fiber-coupling of the two lasers[26] (not shown). The laser powers are adjusted by neutral density (ND) filters. The phonon-side-band (PSB) from the SnV center is detected through a long-pass filter (LPF) using an avalanche photo diode (APD). Confocal scanning to locate the emitters, real-time trace, and PLE measurements are performed with the Qudi python module[38]. SnV emitters and laser conditions are summarized in Supporting Information (Table S1).

The total energy and electronic structure calculation is performed within the scheme of the density functional theory (DFT)[39,40]. The cubic 3x3x3 supercell corresponding to a size of 216 carbon atoms is used and the Γ point is used for the momentum space sampling of the Bloch wave functions. The cutoff energy for the plane-wave basis is set as 64 Ry, and the norm-conserving soft pseudopotential[41] is applied to describe interactions between valence electrons and ions. The local density approximation for the exchange-correlation potential is used by employing the functional form[42] made to match the results of the uniform electron gas[43]. For the charged state, opposite charge uniform background is taken into account to avoid the divergence of the total energy per unit cell. Atomic geometries are optimized according to the force field[44].

**Stability under laser irradiation**

Figure 1(a) shows an atomistic illustration of the SnV center in diamond. A Sn atom is located at an interstitial position and two neighboring carbon sites become vacant. The structure has an inversion symmetry, suppressing the effect of the electrical noise from the environment. Therefore, stable fluorescence with less change in the photon frequency is expected for the SnV center. Four emission lines are observed from the split ground and excited states[14]. In this study, we focus on one of the four lines, called the C-peak, throughout the manuscript.

Figure 1(b) shows a real-time trace of the SnV fluorescence upon simultaneous irradiation of the resonant and non-resonant lasers. The count repeats the bright state with an intensity of ~1.5 kcps and the dark state with the background level. This observation agrees with unstable PLE spectra in previous reports[26,35]. There are two possibilities for instability: change in the photon frequency by spectral diffusion or transition of the charge state. In the following experiments, we reveal that this instability is attributed to the charge state transition of spectrally-stabilized SnV centers with stable photon frequencies under resonant excitation alone.

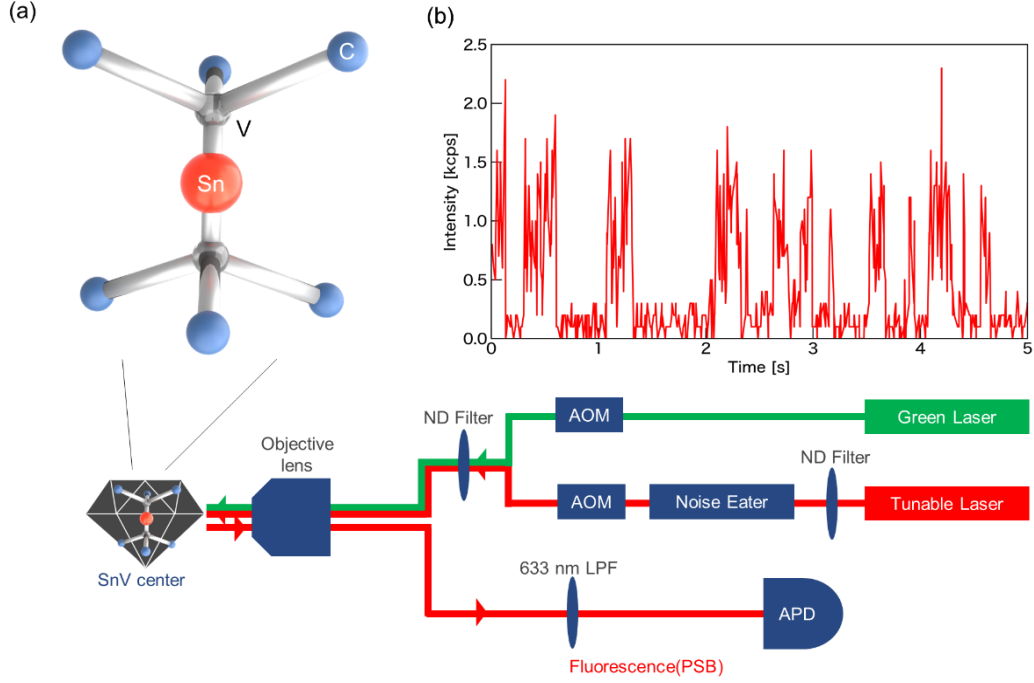

Figure 1. Experimental setup and optical response. (a) Atomic structure of a SnV center and schematic of the home-built confocal system. The blue, red, and grey spheres represent carbon, Sn, and vacancies, respectively. (b) Real-time trace under simultaneous irradiation of resonant and non-resonant lasers at 1 nW and 1 μW, respectively.

We examine the stability of the SnV centers under three conditions of the laser irradiation: weak resonant laser alone (Fig. 2(a)), stronger resonant laser with a green pulse for initialization (Fig. 2(b)), and simultaneous irradiation (Fig. 2(c)). Results of the three types of measurements on one particular SnV center are shown in Fig. 2(d-i). We conduct these measurements for ten SnV centers for statistical analysis. For each emitter, the measurements are carried out in the order of sequence in Fig. 2(a), 2(c), and 2(b). The zero detuning in panels (g-l) corresponds to the center photon frequency in the first scan under only resonant laser.

The first PLE measurements are performed with the resonant laser alone (1 nW), as shown in Fig. 2(a). The emitter shows a narrow PLE peak with a linewidth of 30 MHz (Fig. 2(d)). Figure 2(g) shows the consecutive PLE spectra from the SnV center. The measurements are done in a Boustrophedon manner, in which once the resonant laser is scanned from right to left, then left to right, and so on. Figure 2(j, m) show the summarized histograms of the center photon frequency and linewidth from consecutive PLE scans of the ten SnV center. The different colors correspond to the different SnV centers. The distribution of the center photon frequency has a narrow FWHM of 6.9 MHz. Note that, also including other two sequences, shifts in the center frequency by several tens MHz are only occasionally observed (Fig. 2(j-l)). More details are discussed in Supporting Information (Figs. S1, S2). The histogram of the linewidth shows a narrow distribution with the mean value of ~30 MHz near the transform-limited linewidth, estimated from an excited state lifetime of 5.2 ns[27]. This means that the SnV centers are basically stable over the measurements under the irradiation of the weak resonant laser alone.

Then, we perform the second sequence including a stronger resonant laser (10 nW) for frequency scanning and a non-resonant laser pulse (515 nm, 20 μW, 50 ms) to initialize the charge state of the SnV (Fig. 2(b)). This non-resonant laser is also likely to change the charge environment around the emitter and to induce spectral diffusion. The strong resonant excitation leads to the transition of the SnV to a dark state. In addition to a complete Lorentzian peak, we also observe the termination of the fluorescence near the resonant frequency in another run (Fig. 2(e)). The fluorescence is recovered after the initialization with the green laser at the next scan in the consecutive PLE measurements (Fig. 2(h)). The center photon frequency remains at zero detuning, and no peak appears at another position expected for spectral diffusion or jump. From more 20 times scanning for each SnV centers, we extract the data in which the fluorescence terminates in the previous scan, and then becomes a complete peak in the next scan. This analysis makes it most probable to observe spectral diffusion[36]. One emitter has no data to satisfy this condition, so we analyze the nine emitters for this sequence. We summarize the center photon frequency (Fig. 2(k)) and linewidth (Fig. 2(n)). Even with the fluorescence termination, the histogram of the center photon frequency clearly retains a narrow FWHM of 6.3 MHz similar with the resonant laser alone. This result indicates that our sample formed by the HPHT anneal has less defects resulting in the suppression of spectral diffusion. Note that a SnV center has multiple data at around -50 MHz detuning, but these originate from a previous different sequence (see Supporting Information, Fig. S2). The slight shift of the mean linewidth to a higher value originates from the power broadening with the higher resonant laser power.

As the third sequence, we adopt the simultaneous irradiation of the resonant and non-resonant lasers (Fig. 2(c)), which results in the unstable fluorescence. We find that each PLE spectrum possesses an incomplete peak, repeating the bright fluorescence and background level, as shown in Fig. 2(f). This behavior agrees with the real-time trace in Fig. 1b and previous reports[26,35]. Consequently, the center bright line in the consecutive PLE scans becomes apparently blurred (Fig. 2(i)), suggesting that the SnV centers become unstable during the simultaneous irradiation. We summarize the data showing a linewidth of 20 MHz and higher in curve fitting, comparable to those under weak resonant laser only. Irrespective of the unstable fluorescence, the distribution of the center frequency remains around zero detuning with an FWHM of 17 MHz (Fig. 2(l)), which is smaller than that of the linewidth of standard PLE spectra by only resonant laser in Fig. 2(m). In other words, even if a peak with a linewidth of 30 MHz shifts by ±8.5 MHz, the fluorescence intensity does not go down to the background level.

Therefore, the spectral diffusion does not explain the observed instability under the simultaneous irradiation. Although we do not exclude a possibility that small spectral diffusion occurs, broader distributions of the center photon frequency (Fig. 2(l)) and linewidth (Fig. 2(o)) would mainly come from the incomplete PLE spectra.

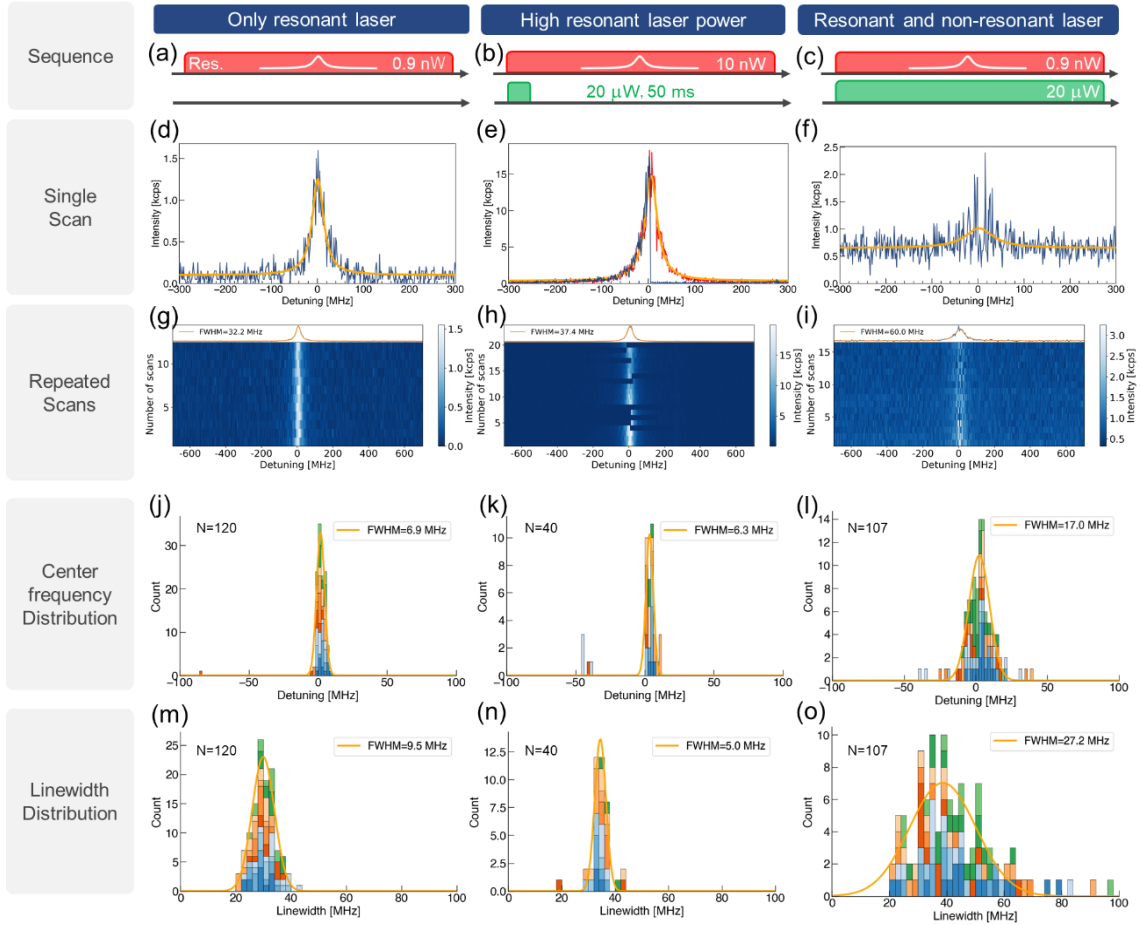

Figure 2. Consecutive PLE measurements under three optical excitation conditions. Sequence of (a) low power resonant laser alone, (b) higher power resonant laser with green charge initialization, and (c) simultaneous irradiation. (d-f) Single PLE spectra and (g-i) consecutive PLE spectra over time of one SnV center. Panel (e) shows a complete peak (red) and a scan with sudden fluorescence termination (blue). Upper panels in (g, i) shows averaged PLE spectra. The data with the maximum intensities in each frequency bin are used to form the peak in panel (h). Histogram of (j-l) center photon frequency and (m-o) linewidth of the PLE spectra of multiple emitters. Each color in the distributions correspond to different SnV centers.

**Dark state transition**

Another possibility of the observed unstable fluorescence is the dark state transition of the SnV center upon the simultaneous irradiation of the resonant and non-resonant lasers. To examine the charge state transition, we investigate time variation of the fluorescence from a SnV center under resonant excitation in the absence and presence of the non-resonant laser (Fig. 3(a,b)). We count the photon number in the phonon-side band and repeat 10,000 times for each condition. Thus, the obtained result is a histogram of the photon number with respect to time. First, as shown in Fig. 3(a), we excite the emitter by resonant laser alone (5 nW) at a fixed laser frequency. The emitter does not show a fluorescence decay on the millisecond-scale, while a decay curve of the fluorescence is observed on the second-scale (inset in Fig. 3(a)), due to the fluorescence termination induced by the change in the charge state of the emitter. We obtain a dark state transition rate of 0.6 Hz from an exponential function fit. Then, we introduce a continuous non-resonant laser (11.5 µW) together with the resonant laser pulse (5 nW), as shown in Fig. 3(b). During the simultaneous irradiation of the resonant and non-resonant lasers, a remarkably rapid decay of the fluorescence is observed with a rate of 1100 Hz. Following the high intensity at the start edge, it decays and reaches a steady state. After ending the resonant pulse, the fluorescence goes down to a weaker level, coming from the diamond surface on the green laser irradiation. Since the decay is triggered by the resonant laser, it is thought to be caused by the dark state transition of the SnV center under the simultaneous irradiation, not spectral diffusion by the non-resonant green laser which is irradiated throughout the 10,000 times repeated runs. Note that the non-resonant excitation also promotes the recovery to the bright state, as shown below. The two processes lead to the fluorescence intensity in the equilibrium between the fluorescence termination and recovery, in agreement with the observation of a GeV center[45].

Next, we perform an experiment using a sequence including multi-pulses of the simultaneous irradiation of the resonant and non-resonant lasers, as shown in Fig. 3(c). After the first pulse of the non-resonant laser for the charge initialization, we apply a 5 ms resonant laser pulse to check the standard fluorescence intensity of this emitter (~1500 cps) as a reference. Then, we apply a 0.1 ms simultaneous irradiation of the resonant and non-resonant lasers, and we again use the resonant laser alone to readout the fluorescence intensity of the emitter (~1200 cps), which is slightly lower compared to the standard fluorescence intensity. We repeat the simultaneous irradiation-readout pulse block once more and observe a further decrease in the fluorescence intensity to ~1000 cps. Finally, we apply a non-resonant laser pulse and subsequently readout the fluorescence intensity, and find that the fluorescence intensity recovers to the standard value. The stepwise reduction in the fluorescence intensity after each pulse of the simultaneous irradiation indicates the controllability of this process, not due to stochastic spectral diffusion.

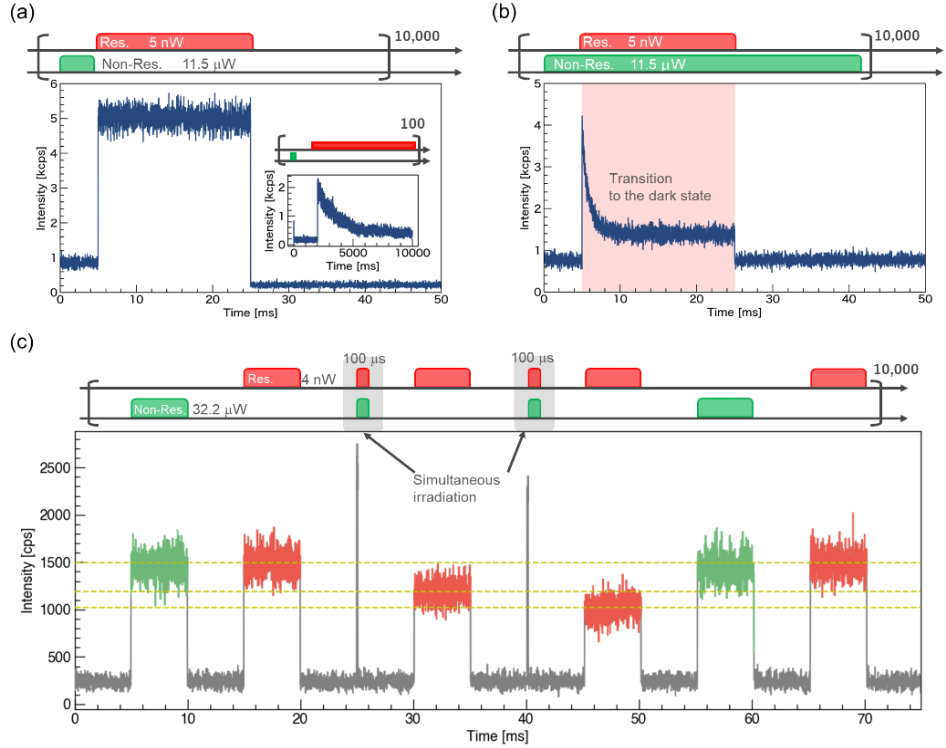

Figure 3. Pulse excitation of SnV center. (a) Fluorescence response under resonant excitation. Repetition: 10,000 times. The inset shows a decay during a longer time scale. An 8 s resonant pulse (4 nW) is irradiated. Repetition: 100 times. (b) Fluorescence response under simultaneous irradiation of non-resonant (11.5 μW) and resonant (5 nW) lasers. Repetition: 10,000 times. (c) Multi-pulses of simultaneous excitation. Two simultaneous excitations (0.1 ms for each) and subsequent readout resonant pulses are irradiated. In the pulse sequence in the upper panel, the pulse widths of the simultaneous irradiation are illustrated longer than the actual duration for the sake of clarity. Repetition: 10,000 times. The dashed lines indicate the mean fluorescence intensities under the resonant excitation.

**Laser power dependence of decay rate**

Then, we investigate the decay rate under different laser powers to examine a possible path to the dark state upon the simultaneous irradiation. Figure 4(a) shows the decay curves at various resonant laser powers at a fixed non-resonant laser power of 20 μW. The decay becomes faster as increasing the resonant laser power, and it is well fitted linearly with a slope of 301 Hz/nW, as shown in Fig. 4(b). This indicates that the transition process to the dark state involves one resonant photon. Importantly, here, we clearly see that the resonant laser power influences the decay rate although the emitter does not go to the dark state on this time-scale by the resonant laser alone as shown in Fig. 3(a). Therefore, it is thought that the resonant excitation works for efficiently making the SnV center in the excited state of the -1 charged state, and then, the SnV in the excited state goes to the dark state with addition of the non-resonant laser.

The dependence of the decay rate on the non-resonant laser power gives an insight to the transition process to the dark state (Fig. 4(c)). The decay rate almost linearly increases up to 30 μW (Fig. 4(d)). In contrast, the error bar at 40 μW becomes significantly larger and the rate drops at 50 μW. These higher powers of the non-resonant laser could cause large spectral diffusion and unstable measurements. Indeed, the simultaneous irradiation with 50 μW green laser resulted in a broad linewidth of 580 MHz[26]. Thus, we analyze the data up to 30 μW in Fig. 4(d). The linear behavior at the low power range suggests involvement of one non-resonant photon to the dark state transition. The linearity at the low powers and instability at the high powers are also observed at a different laser condition (Supporting Information, Fig. S3).

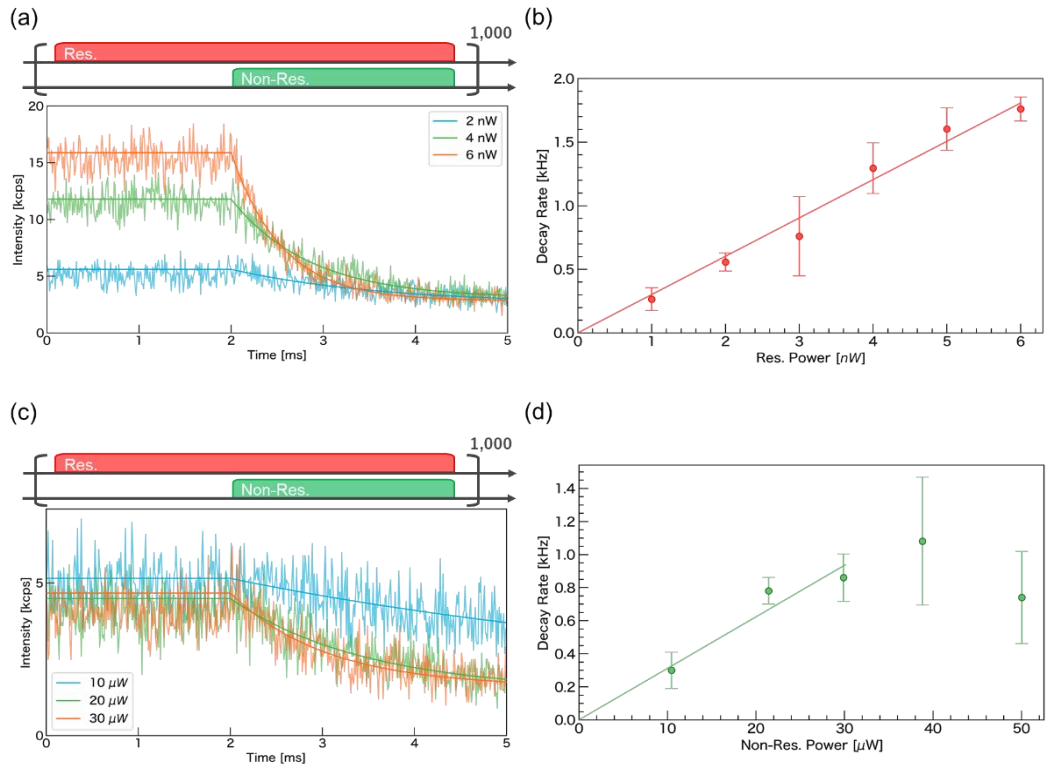

Figure 4. Decay rate. (a) Time variation and (b) decay rate as a function of the resonant laser power. (c) Time variation and (d) decay rate as a function of the non-resonant laser power. The solid lines between 0 and 2 ms in panels (a, c) correspond to the values at 2 ms of each decay curve fitting. The decay rates in panels (b, d) are obtained from the curve fitting without considering initialization rates by the non-resonant laser because of much lower values as seen in Fig. 5. Each sequence includes an initialization pulse (omitted in the illustration of the sequences).

**Recovery rate from dark state**

We observe the recovery of the fluorescence with the non-resonant green laser. Figure 5(a) shows the sequence and observed fluorescence from a SnV center. A 16 times-repeated block including the non-resonant and resonant pulses works as the gradual charge stabilization and photon counting from the SnV center, and the simultaneous irradiation is intended to change to the dark state. The fluorescence gradually increases as increasing the number of the non-resonant pulses and is saturated at around the 6th pulse. Figure 5(b) shows the recovery rate as a function of the non-resonant laser power. Note that, for these measurements, the initial state is not the complete dark state, but the steady state upon the simultaneous irradiation. As expected, basically, a higher laser power leads to a higher recovery rate to the bright state. The recovery rate has a linear dependence on the laser power, indicating that one photon in the non-resonant laser involves this recovery process[35]. The observation here mentions that the recovery occurs in a controlled manner, again evidencing the dark state transition upon the simultaneous irradiation.

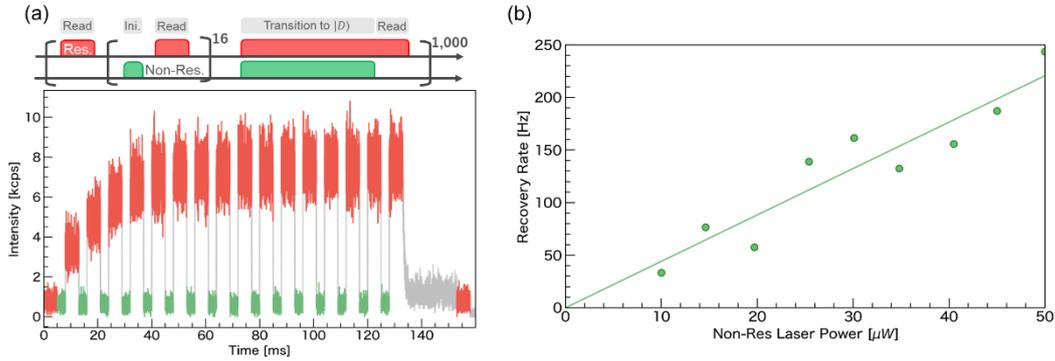

Figure 5. Recovery from the dark state. (a) Pulse sequence and observed fluorescence. (b) Recovery rate. $|D\rangle$ in the sequence denotes the dark state.

**Model of charge state transition**

Finally, we consider the mechanism of the charge state transition of the SnV center upon the simultaneous irradiation. The energy levels in Fig. 6 are estimated by first-principle calculations. (i) First, in the ground state of the SnV⁻ center, one of the $e_u$ electrons is efficiently excited to the $e_g$ level by absorbing one resonant photon. (ii) So, the SnV⁻ center becomes in the excited state. (iii) There are two possibilities of the dark state: SnV⁰ or SnV²⁻. Considering the underestimation of the band gap of diamond in our calculation, the energy of the 515 nm laser could not be sufficient to pump up an electron from the $e_g$ levels in an excited SnV⁻ to the conduction band of diamond. Furthermore, the calculation shows that there is no energy level near the bottom of the conduction band which accepts electrons from the $e_g$ levels. Thus, the SnV$^-$ does not likely go to neutral SnV$^0$ under this lase condition. Instead, an electron is thought to be excited from the valence band to the half-empty $e_u$ level of the excited SnV center, with a resonant photon (red dashed arrow) or a non-resonant green photon (green solid arrow). Indeed, in calculation, we find several energy levels in the diamond valence band from which optical excitation to the $e_u$ level is feasible. These energy levels are denoted as the blue dashed line. Experimentally, while the decay rate is as low as 0.6 Hz under only the resonant excitation (4 nW, inset of Fig. 3(a)), it significantly increases to ~1100 Hz under the simultaneous irradiation even at a similar weak resonant laser power (5 nW). This indicates that the non-resonant green laser with a higher power of 11.5 μW included in the simultaneous irradiation contributes to this electron excitation. Thus, the dark state transition occurs quickly upon the simultaneous irradiation. Note that it is thought that the two lasers are not necessarily irradiated at the same time, but consecutive irradiation (resonant, then non-resonant excitation) with an intermediate time less than the excited state lifetime of ~5 ns[27] should work for the dark state transition. (iv) When both the $e_u$ and $e_g$ levels are fully occupied by electrons, the SnV center takes the -2 charge state[35]. Then, the dark SnV$^{2-}$ goes back to the bright SnV$^-$ with the non-resonant laser. Our calculation suggests that the photon energy at 515 nm is enough to pump up an electron from the SnV$^{2-}$ center to the conduction band, which is consistent with previous calculation[34]. As another possibility, according to the model[35], a hole is generated from an optically excited defect, e.g. a divacancy, and recombines with an electron in the SnV$^{2-}$ center, leading to the recovery to the bright state.

It is worth noting that as another path to the dark state, there is the possibility of the direct transition from the ground state of the SnV$^-$ center, as mentioned in a GeV center[45]. However, we exclude this possibility as the main route to the dark state because, as shown in Fig. 4(b), the dark state transition becomes faster even for the lower population in the ground state on the higher resonant laser power (higher population in the excited state as evidenced by the increase in the fluorescence intensity in Fig. 4(a)). Thus, it is crucial to pump up to the excited state once by the resonant laser for the dark state transition.

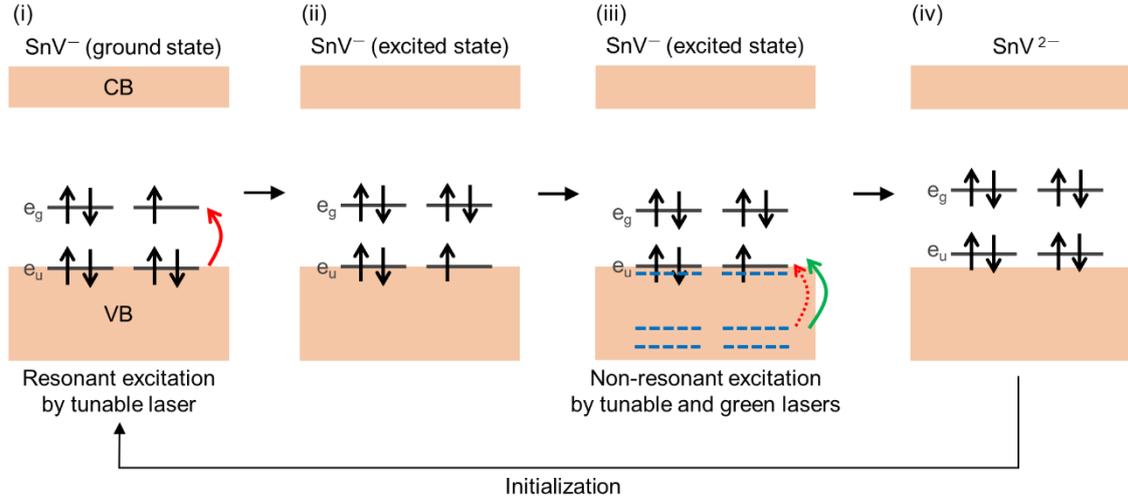

Figure 6. Model of charge state transition. VB and CB denote the valence band and conduction band of diamond, respectively. The $e_g$ and $e_u$ levels are degenerate states of the SnV center. Blue dashed lines denote states in the diamond valence band, located at around 0.2, 1.6, and 2.1 eV from the $e_u$ levels. The wavelength of the tunable laser is set to the resonance for the SnV$^-$ excitation shown in panel (i), making it non-resonant between the $e_u$ level and blue lines in process (iii).

**Conclusion**

We investigated the spectral stability and charge state transition of the SnV centers in diamond. The resonant frequencies of the SnV centers were stable over time under resonant excitation even with the initialization pulse of the non-resonant laser. In contrast, the PLE spectra became unstable under simultaneous irradiation of the resonant and non-resonant lasers, showing the repetition of the termination and recovery of the fluorescence. We found that this phenomenon originates from the dark state transition of the SnV centers from the bright state, not spectral diffusion. The SnV center efficiently transitions to the excited state under resonant excitation, and then the transition to the dark SnV$^{2-}$ state occurs by pumping an electron from the valence band of diamond under non-resonant laser. Although the non-resonant laser also works for initialization to the bright SnV$^-$ state, its contribution to the dark state transition is much larger under the simultaneous irradiation. Consequently, the SnV center goes to the dark state. The SnV$^{2-}$ center goes back to the bright SnV$^-$ center by non-resonant laser irradiation in a controlled manner. This work will contribute to achieving the fabrication of stable SnV centers and give an insight into their charge state control, which are essential to construct quantum network nodes.

**Acknowledgments**

We thank Y. Narita for experimental support. This work is supported by JSPS KAKENHI Grant Numbers JP22H04962, the MEXT Quantum Leap Flagship Program (MEXT Q-LEAP) Grant Number JPMXS0118067395, and JST Moonshot R&D Grant Number JPMJMS2062.

# Supporting Information: Charge state transition of spectrally stabilized tin-vacancy centers in diamond

Keita Ikeda[1], Yiyang Chen[1], Peng Wang[1], Yoshiyuki Miyamoto[2], Takashi Taniguchi[3], Shinobu Onoda[4], Mutsuko Hatano[1], Takayuki Iwasaki[1,]

[1]Department of Electrical and Electronic Engineering, School of Engineering, Institute of Science Tokyo, Meguro, 152-8552 Tokyo, Japan
[2]Research Center for Computational Design of Advanced Functional Materials, National Institute of Advanced Industrial Science and Technology, Tsukuba, Ibaraki 305-8568, Japan
[3]Research Center for Materials Nanoarchitectonics, National Institute for Materials Science, 305-0044 Tsukuba, Japan
[4]Takasaki Advanced Radiation Research Institute, National Institutes for Quantum Science and Technology, 1233 Watanuki, Takasaki, 370-1292 Gunma, Japan

## 1. SnV emitters and measurement conditions.

Table SI summarizes the emitter number and laser conditions used in each figure.

Table S1. The SnV center and laser powers in each figure.

| Figure number | Emitter | Resonant power | Non-resonant power |
|---|---|---|---|
| 1(b) | #1 | 1 nW | 1 μW |
| 2 | #2~#11 | 0.9, 10 nW | 0, 20 μW |
| 3(a) | #12 | 5 nW | 11.5 μW |
| 3(a), inset | #12 | 4 nW | 15.4 μW |
| 3(b) | #12 | 5 nW | 11.5 μW |
| 3(c) | #13 | 4 nW | 32.2 μW |
| 4(a) | #14 | 2, 4, 6 nW | 20 μW |
| 4(b) | #14 | 1 – 6 nW | 20 μW |
| 4(c) | #14 | 2 nW | 10, 20, 30 μW |
| 4(d) | #14 | 2 nW | 10 – 50 μW |
| 5(a) | #14 | 5 nW | 30.1 μW |
| 5(b) | #14 | 5 nW | 10 – 50 μW |
| S1 | #6 | 0.9 nW | - |
| S2 | #5 | 0.9, 10 nW | 0, 20 μW |
| S3 | #14 | 5 nW | 10 – 50 μW |

## 2. Instability of the SnV emission in consecutive PLE scans

The SnV centers studied in this study basically show stable PLE scans, but center frequency shifts by several tens MHz are only occasionally observed. Figure S1(a) shows consecutive PLE scans of an emitter under only weak resonant laser (0.9 nW). Although we observe two bright lines, corresponding to two SnV centers with similar photon frequencies, a shift of the center frequency is seen for one emitter. It might originate from spectral diffusion, but we do not exclude the possibility that the drift of the laser spot occurs because both bright lines almost disappear at the next scan.

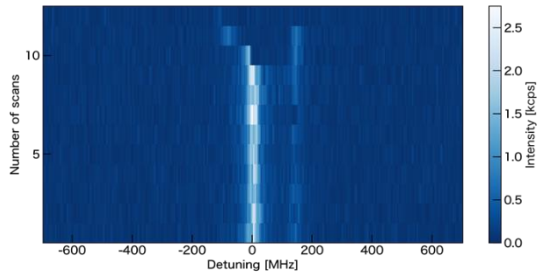

Figure S1. PLE scans under low power resonant laser alone (0.9 nW).

PLE scans on another emitter are shown in Fig. S2. The measurements are carried out in the order of sequence: weak resonant laser alone, simultaneous irradiation, and stronger resonant laser with non-resonant green pulse for charge initialization. The emitter shows stable fluorescence under weak resonant alone. In contrast, under the simultaneous irradiation, the center frequency is shifted by ~40 MHz at the 14th scan. This shift continues in the next sequence with the stronger resonant laser, while the center frequency remains stable during this sequence.

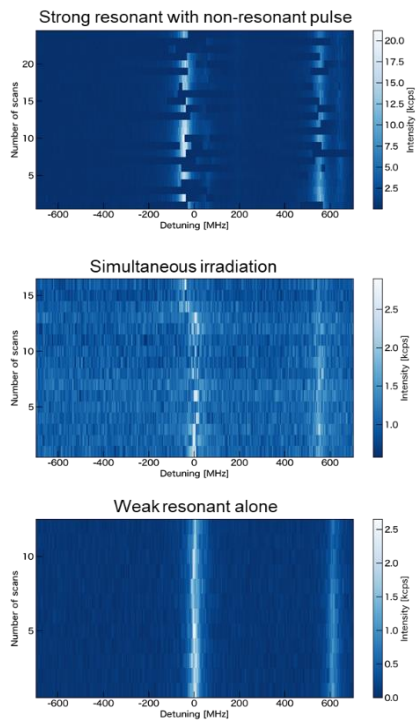

Figure S2. PLE scans under three different laser conditions on another emitter. The zero detuning in panels corresponds to the center photon frequency in the first scan under weak resonant laser alone.

## 3. Non-resonant laser power dependence on decay rate

Figure S3 shows a decay rate depending on the non-resonant laser power with a different resonant laser power from Fig. 4(d) in the main text. We observe similar behaviors such as linear dependence up to 30 μW, a large error bar at 40 μW, and reduction in the decay rate at 50 μW.

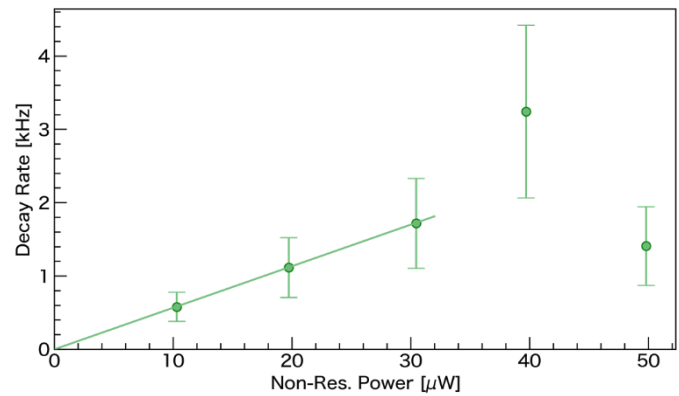

Figure S3. Decay rate as a function of the non-resonant laser power. The resonant laser power is 5 nW.